\documentclass[12pt]{iopart}

\usepackage{cite}
\usepackage{iopams}
\usepackage{amssymb}
\usepackage{amscd}
\usepackage{wasysym}

    \usepackage[dvips]{graphicx}
    \DeclareGraphicsExtensions{.eps}

\newcommand{\be}{\begin{equation}}
\newcommand{\ee}{\end{equation}}
\newcommand{\bea}{\begin{eqnarray}}
\newcommand{\eea}{\end{eqnarray}}
\newcommand{\nn}{\nonumber}
\newcommand{\ket}[1]{|#1\rangle}
\newcommand{\bra}[1]{\left\langle#1\right|}

\newcommand{\ri}{{\rm i}}
\newcommand{\re}{{\rm e}}
\newcommand{\rd}{{\rm d}}

\begin{document}

\title{Photon scattering by a three-level emitter in a one-dimensional waveguide}

\author{D Witthaut${}^{1,2}$ and A S S\o{}rensen$^{1}$}

\address{${}^1$QUANTOP, The Niels Bohr Institute, 
               University of Copenhagen, DK--2100 Copenhagen, Denmark\\
	       ${}^2$Max-Planck-Institute for Dynamics and Self-Organization,
	       D--37073 G\"ottingen, Germany
}

\ead{dirk.witthaut@nbi.dk}

\date{\today }

\begin{abstract}
We discuss the scattering of photons from a three-level emitter in a
one-dimensional waveguide, where the transport is governed by
the interference of spontaneously emitted and directly transmitted 
waves. The scattering problem is solved in closed form for different 
level structures. Several possible applications are discussed: 
The state of the emitter can be switched deterministically by Raman 
scattering, thus enabling applications in quantum computing such as 
a single photon transistor.
An array of emitters gives rise to a photonic band gap structure, 
which can be tuned by a classical driving laser. A disordered
array leads to Anderson localization of photons, where the 
localization length can again be controlled by an external driving. 
\end{abstract}


\maketitle


\section{Introduction}

The effects of spontaneous emission have been extensively 
explored in two limits. In the weak coupling limit, spontaneous
emission is essentially viewed as irreversible loss and a
source of decoherence. On the contrary, excitations are 
periodically exchanged between an emitter and the modes 
of a cavity in the regime of strong coupling.
A different regime is explored in one-dimensional waveguides,
where photon scattering is goverend by the interference of 
the absorbed, reemitted and the directly transmitted wave. 
For a simple two-level emitter this leads to a complete reflection 
if the photon is resonant with the atomic transition 
\cite{Shen05,Pino08,Zumo08}.
In the present paper we discuss photon scattering from a
three-level emitter, which can lead to even richer structures
in reflection and transmission. 

In order to reach the one-dimensional regime,  the coupling to the 
waveguide has to be strong compared to transversal losses. The 
strong coupling regime has been first realized in a high--Q resonator 
for optical or microwave photons (see, e.g. \cite{Thom92,Brun96}).
Later, strong coupling was also demonstrated in nanoscale integrated 
devices using photonic crystal microcavities \cite{Reit04,Yosh04}.
In such a cavity, however, the emitter couples only to a discrete set 
of modes and not to a one-dimensional continuum as in a waveguide.
Great advances to reach the regime of strong coupling also in this 
situation have been reported only recently using several different
systems such as photonic crystal waveguides \cite{Fara07}, tapered 
optical fibers \cite{Daya08} or hollow core fibers \cite{Bajc09}. 
Another opportunity is to exploit the strong localization of 
surface plasmon polariton (SPP) modes along a metallic nanowire
\cite{Chan06,Chan07,Akim07,Haka09}. Strong coupling seem to be 
feasible even in free space if the photons are focussed 
appropriately \cite{Hwan09}. Some of these different setups 
have been sketeched in figure \ref{fig-experiments}. Thus the 
peculiar effects discussed in the present paper may
become experimentally accessible in the near future.

In the present paper we solve the scattering problem for photons
in one-dimensional waveguides coupled to a single three-level
emitter in various configurations (cf. figure \ref{fig-levels3}).
It is shown that a waveguide coupled to a $V$-type emitter or a
driven $\Lambda$-type emitter shows a characteristic EIT-like 
transmission and reflection spectrum.
Raman scattering occurs for a $\Lambda$-type emitter.
Most remarkably, the Raman scattering amplitude can be turned to 
one on resonance, so that the photon is deterministically transferred 
to a sideband, flipping the quantum state of the emitter. 
These features allow for a variety of applications in quantum optics
and quantum information. A possible realization of a single photon 
transistor from a driven emitter is discussed in 
section \ref{sec-transistor}. The transmission through a regular array
of emitters is discussed in section \ref{sec-bloch}. It is shown that
the resulting photonic Bloch bands are widely tunable with an
external classical control field. Disorder in the emitter positions
leads to Anderson localization of the photonic eigenstates, where 
the localization length can also be controlled by the external field
as shown in section \ref{sec-anderson}.

\section{Single photons in nanoscale waveguides}
\label{sec-setup}

\begin{figure}[tb]
\centering
\includegraphics[width=13cm, angle=0]{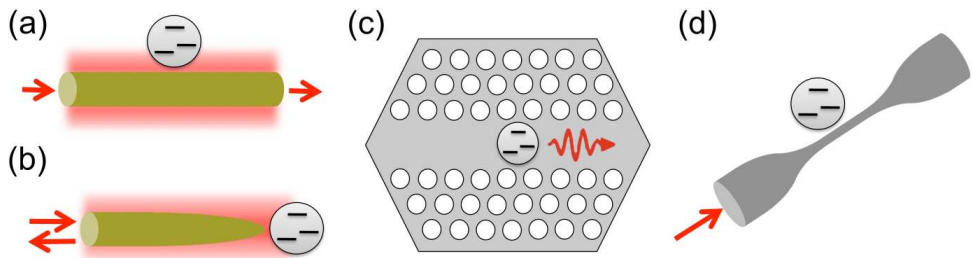}
\caption{\label{fig-experiments}
Possible experimental realizations of a one-dimensional waveguide
strongly coupled to a single emitter: a surface plasmon polariton 
mode on a metallic nanowire (a) and on a metallic nanotip (b),
a guided mode in a photonic crystal waveguide (c) or a
tapered optical fiber (d).
}
\end{figure}

In this section we solve the scattering problem for a photon in a
quasi one-dimensional geometry as sketched in figure \ref{fig-experiments}.
The dynamics is governed by two effects which are usually not present
in free space, the strong coupling to a single emitter and the interference 
of spontaneously emitted waves. In particular, the dynamics in such a 
one-dimensional waveguide is modelled by the Hamiltonian \cite{Shen05,Shen07,Chan07b}
\be
  \hat H_{\rm tot} = 
  \hat H_{\rm free} + \hat H_{\rm atom} + \hat H_{\rm int}.
  \label{eqn-ham0}
\ee
The free propagation of the photons is described by the Hamiltonian
\be
  \hat H_{\rm free} = - \ri c  \int \rd x \,
    \hat a_R^\dagger(x) \frac{\partial}{\partial x } \hat a_R(x)
   - \hat a_L^\dagger(x) \frac{\partial}{\partial x } \hat a_L(x).
   \label{eqn-ham-free}
\ee
Here, $\hat a_R(x)$ and $\hat a_R(x)$ denote the annihilation 
operators of a right- or left-moving photon at position $x$, respectively.
The separation into left- and right-moving modes is possible as we
are dealing with optical photons with a frequency width much smaller
than the mean frequency $\omega_0$. Thus one can safely decompose
the field into two distinct contributions with positive and negative
wavenumbers, corresponding to right- and left-moving modes.
The interaction Hamiltonian is local at the position of the emitter $x_0$,
and is given by
\be
  \hat H_{\rm int} = \bar g \int \rd x  \, \delta(x-x_0) 
   \hat S_+ (\hat a_R(x)  + \hat a_L(x) )
    + \mbox{h. c.},
   \label{eqn-ham-int}
\ee
where $\hat S_-$ and $\hat S_+ = \hat S_-^\dagger$ denote
atomic lowering and raising operators, respectively.
We will consider different structures of the emitter described
by different Hamiltonians $\hat H_{\rm atom}$, which will be
specified for the respective situations below.
In the following we set $\hbar = 1$, thus measuring all energies
in frequency units.

The dynamics is simplified by setting the coordinates such that
$x_0 = 0$ and introducing the (anti)-symmetric modes
\bea
  {\hat a}_e(x) &=& \frac{1}{\sqrt{2}} 
       \left( {\hat a}_R(x) + {\hat a}_L(-x)   \right), \nn \\
  {\hat a}_o(x) &=& \frac{1}{\sqrt{2}} 
       \left( {\hat a}_R(x) - {\hat a}_L(-x)   \right),
       \label{eqn-lrmodes}
\eea
so that the Hamiltonian for the symmetric mode reads
\bea
  \hat H_{\rm free} &=& - \ri c \int \rd x \,
   \hat a_e^\dagger(x) \frac{\partial}{\partial x } \hat a_e(x) \nn \\
  \hat H_{\rm int} &=& g \int \rd x \, \delta(x) 
   \left( \hat S_-  \hat a_e^\dagger(x) +  \hat S_+  \hat a_e(x) \right),
   \label{eqn-ham1}
\eea
with $g =  \sqrt{2} \bar g$
while the antisymmetric-mode $\hat a_o$ does not couple to the emitter at all.
In the following we consider only the symmetric mode and thus drop 
the  index $e$. 

Then it is easy to show that the rate of spontaneous emission into the
one-dimensional waveguide is given by $\Gamma = g^2/c$. Below
we mostly use $\Gamma$ to characterize the interaction strength.
Spontaneous emission to other modes out of the one-dimensional
waveguide is modelled by attributing an imaginary part $- \ri \gamma/2$
to the energies of the excited levels in $\hat H_{\rm atom}$ in the
spirit of the quantum jump picture \cite{Dali92,Carm93}.
 
A more efficient coupling may be realized by placing the emitter at 
one end of a semi-inifinite waveguide instead of side-to-side. This 
was shown in particular for plasmonic waveguides \cite{Chan07},
where the electric field strength around a nanotip is significantly incresed
(cf. figure \ref{fig-experiments} (b)) The propagation of the right- 
and left-going  modes is then restricted to $x<0$.
In this case we introduce the mode function
\be
  \hat a_e(x) = \left\{ \begin{array}{c l}
    \hat a_R(x) & \mbox{for} \; x < 0 \\
    \hat a_L(-x) & \mbox{for} \; x > 0, \\
    \end{array} \right.
\ee
such that $x<0$ describes the incoming and $x>0$ the reflected
photons.
The original Hamiltonian (\ref{eqn-ham0}) then also assumes the
form (\ref{eqn-ham1}), however with $g = \bar g$.

The basic scattering problem for a simple two-level emitter
has been solved by Shen and Fan \cite{Shen05}. 
In this case we have $\hat H_{\rm atom} =  \omega_0 \sigma_z/2$
and $\hat S_+ = \sigma_+$, where $\sigma_j$ denote the respective
Pauli matrices. 
The solution starts from an eigenstate of the full Hamiltonian which
can be written as 
\be
  \ket{\Psi} = \int \rd x \, f_g(x) \hat a^\dagger(x) \ket{g,\emptyset} 
      + f_e \ket{e,\emptyset},
  \label{eqn-1phot-in}
\ee
where $\ket{g}$ and $\ket{e}$ are the ground and excited state of
the emitter, respectively. Here and in the following $\ket{\emptyset}$ 
denotes the state of zero photons, i.e. the empty waveguide. The 
mode function is discontinuous due to the $\delta$-function in the
interaction Hamiltonian,
\be
   f_g(x) = \left\{ \begin{array}{l c l}
       f_{\rm in}(x) & \; \mbox{for} \;  & x>0 \\
       f_{\rm out}(x) & & x>0. \\ 
     \end{array} \right.
\ee
Using the Lippmann Schwinger formalism one can then rigorously 
show that an input state given by the mode function $f_{\rm in}(x)$
is scattered to an output state given by the mode function 
$f_{\rm out}(x)$ \cite{Shen07}. In particular one finds that the transmission 
amplitude for a monochromatic input state with wavenumber $k$ 
is given by 
\be
  t_k = \frac{ck - \omega_0 + \ri ( \gamma - \Gamma)/2}{
                   ck - \omega_0 + \ri ( \gamma + \Gamma)/2} \, .
  \label{eqn-tk}
\ee

This procedure is readily generalized to three-level emitters, but 
some attention has to be paid in the case of two groundstates, i.e. 
a $\Lambda$-type atom as shown in figure \ref{fig-levels3} (B).

\section{Scattering by a three-level atom}
\label{sec-scat}

\begin{figure}[tb]
\centering
\includegraphics[width=10cm, angle=0]{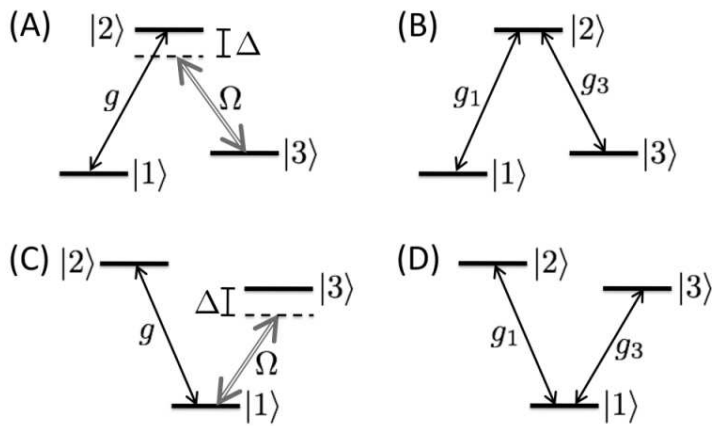}
\caption{\label{fig-levels3}
The different atomic level schemes considered in the present paper:
(A) Electromagnetically induced transparency in a 
    driven $\Lambda$-system,
(B) Raman scattering in a $\Lambda$-system,
(C) Raman scattering in a driven $V$-system,
(D) Electromagnetically induced transparency in a $V$-system.    
}
\end{figure}

In this section we solve the scattering problem for a single photon
and a three-level emitter with three internal states. The different 
possibilities for the coupling and a classical driving field are sketched
in figure \ref{fig-levels3}.

\subsection{Electromagnetically induced transparency in a 
    driven $\Lambda$-system}

We start with the driven $\Lambda$ level scheme shown in 
figure \ref{fig-levels3} (A),
where the excited atomic state $\ket{2}$ is coupled to another level
$\ket{3}$ by a classical laser beam with Rabi frequency $\Omega$ and
detuning $\Delta$. Within the rotating wave approximation, the atomic
Hamiltonian is given by
\bea \fl \qquad
  H_{\rm atom} =
        (E_2 - \ri \gamma_2/2) \ket{2}\bra{2} 
          + (E_2-\Delta- \ri \gamma_3/2 ) \ket{3}\bra{3} 
     + \frac{\Omega}{2} \left( \ket{3}\bra{2} + \ket{2}\bra{3} \right),
\eea
where we have set the energy scale such that the energy of level $\ket{1}$
is zero. The interaction Hamiltonian is given by equation (\ref{eqn-ham1}) with
$S_+ = \ket{2}\bra{1}$. 

The eigenstates of the full Hamiltonian (\ref{eqn-ham0}) are then given by
\be
  \ket{E} = \int \rd x f_1(x) \hat a^\dagger(x) \ket{\emptyset,1}
     + f_2 \ket{\emptyset,2} + f_3 \ket{\emptyset,3}
\ee
with eigenenergy $E = ck$. The coefficients are easily found to be
\bea
  f_1(x) &=& \frac{1}{\sqrt{2\pi}} \left( \Theta(-x) + t_k \Theta(+x) \right) 
              \re^{\ri kx} \nn \\
  f_2 &=& \frac{1}{\sqrt{2\pi}}  \frac{\ri c(t_k-1)}{\sqrt{c\Gamma}} \\  
  f_3 &=& \frac{1}{\sqrt{2\pi}}  \frac{\ri c\Omega(t_k-1)}{2(ck+\Delta)}  \nn,
\eea
where the transmission coefficient is given by
\be
  t_k = \frac{[ck-(E_2-\Delta-\ri \gamma_3/2)]
                   [ck-(E_2-\ri \gamma_2/2)- \ri \Gamma/2] - \Omega^2/4}{
                   [ck-(E_2-\Delta-\ri \gamma_3/2)]
                   [ck-(E_2- \ri \gamma_2/2)+ \ri \Gamma/2] - \Omega^2/4 }
  \label{eqn-tk-casea}
\ee
and $\Theta(x)$ denotes the Heaviside step function.
An incoming photon with wavenumber $k$ in the $e$-mode thus 
experiences a phase shift given by $t_k$ when it crosses the atom. 
The modulus of $t_k$ is smaller than one for $\gamma_2 > 0$
or $\gamma_3 >0$ which
describes transversal losses, i.e. the scattering of the photon out of 
the waveguide. In this case the emitter jumps 
incoherently to one of the ground levels and the photon is lost.

\begin{figure}[tb]
\centering
\includegraphics[width=13cm, angle=0]{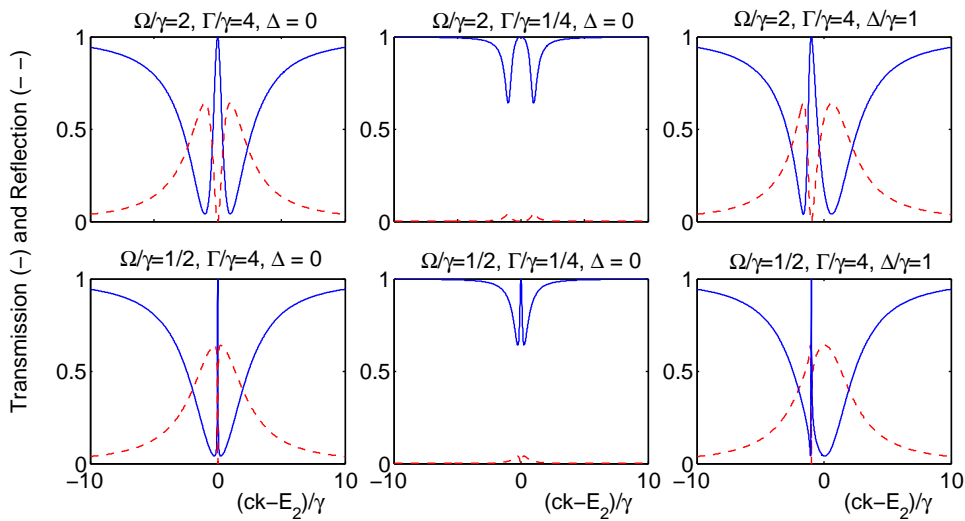}
\caption{\label{fig-eit1}
Transmission (solid line) and Reflection (dashed line) spectrum of the 
driven $\Lambda$-type atom for different values of the parameters 
$\Omega$, $\Gamma$ and $\Delta$.
}
\end{figure}

Let us first consider the case where level $\ket{3}$ is metastable,
i.e. $\gamma_3 = 0$. This situation is realized, for example, when
the levels $\ket{1}$ and $\ket{3}$ are two hyperfine levels in the
ground state manifold coupled by a radio-frequency driving field.
Going back to left- and right-moving modes using equation 
(\ref{eqn-lrmodes}), one finds that the transmission and 
reflection amplitudes of the waveguide are given by
\bea
  \tilde t_k = \frac{t_k + 1}{2} \qquad \mbox{and} \qquad
  \tilde r_k = \frac{t_k - 1}{2}.
  \label{eqn-rt-tilde}
\eea
The resulting transmission spectrum is depicted in figure \ref{fig-eit1}
for different values of the parameters $\Omega$,  $\Delta$ and $\Gamma$,
showing the familiar EIT transmission spectrum \cite{Flei05}.
The atom becomes 
fully transparent on the photon resonance, i.e. when the wavenumber 
is given by
\be
  ck = E_2 - \Delta.
\ee
The width of this central transpareny window is only given by the strength
of the driving field $\Omega$. Furthermore the system is transparent for
$(ck - E_2) \rightarrow \pm \infty$, where the incoming photon is far off-resonant
and thus does not interact with the atom. 
In between we find dips of the transmission, whose width and depth increases
with the coupling strength $\Gamma$. These are mainly due to reflection if 
$\Gamma > \gamma$ and due to losses otherwise. Complete reflection is 
only possible if the losses vanish exactly ($\gamma = 0$) and 
\be
  c k =  E_2 - \frac{\Delta}{2} \pm \frac{\Omega_{}\rm eff}{2} ,
  \label{eqn-casea-reflect}
\ee
where $\Omega_{\rm eff} = (\Omega^2 + \Delta^2)^{1/2}$ denotes the 
effective Rabi frequency.
The complete spectrum is symmetric only for $\Delta = 0$ and asymmetric
otherwise.

\begin{figure}[tb]
\centering
\includegraphics[width=14cm, angle=0]{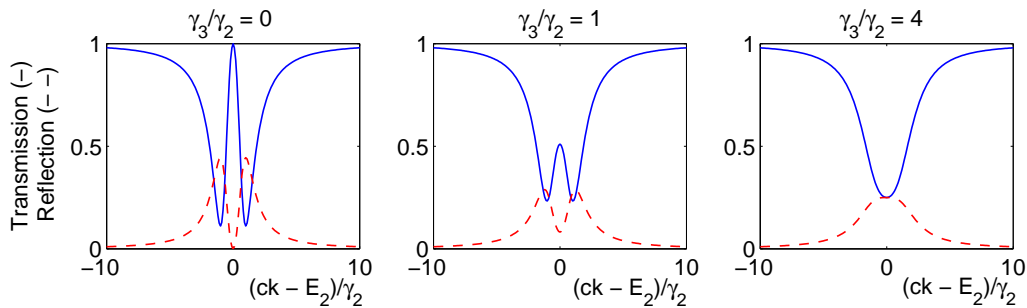}
\caption{\label{fig-noeit1}
Transmission (solid line) and Reflection (dashed line) spectrum of the 
driven $\Lambda$-type atom for different values of $\gamma_3$. The 
remaining parameters are chosen as $\Omega/\gamma_2 = 2$, 
$\Gamma/\gamma_2 = 2$  and $\Delta = 0$.
}
\end{figure}

The waveguide can be fully transparent only if level $\ket{3}$ is 
metastable. For $\gamma_3 > 0$ there will always be losses such
that $|\tilde t_k|^2 + |\tilde r_k|^2 < 1$.
Figure \ref{fig-noeit1} shows how the transmission spectrum changes
when $\gamma_3$ is increased: The transmission on resonance
decreases until finally  the EIT transparency window vanishes completely.

\subsection{A $\Lambda$-system with two coupling transitions}

The $\Lambda$-system shown in figure \ref{fig-levels3} (B), where both 
ground states couple to the waveguide mode is described by the atomic 
Hamiltonian
\bea
  H_{\rm atom} =  E_1 \ket{1}\bra{1} + E_3 \ket{3}\bra{3} 
  + (E_2 - \ri \gamma/2) \ket{2}\bra{2} \nn
\eea
and the interaction Hamiltonian
\bea
  H_{\rm int} = \int \rd x \, \delta(x) \,  
   \big[ g_1  \hat a^\dagger(x) \ket{1}\bra{2} 
        + g_3 \hat a^\dagger(x) \ket{3}\bra{2} 
          + \mbox{h.c.} \big]   \, .
\eea	
Now we have to take into account that the atom can be in state
$\ket{1}$ and $\ket{3}$ when a photon is present. An arbitrary state
with one excitation can thus be written as
\be
  \ket{\Psi} = \int \rd x \left[ f_1(x) \hat a^\dagger(x) \ket{\emptyset,1} 
         + f_3(x) \hat a^\dagger(x) \ket{\emptyset,3}  \right]
          + f_2 \ket{\emptyset,2}. 
\ee
The eigenstates of the full Hamiltonian are found by substituting this ansatz
into the time-independent Sch\"odinger equation. Then one finds two types of
solution:

The continous solution is characterized by a destructive interference
of the functions $f_1(x)$ and $f_3(x)$ at the position of the atom so
that this mode does not interact. The solution is thus given by
\bea
  f_1(x) &=& \mathcal{N} g_3 \re^{\ri k_1 x}, \nn \\
  f_3(x) &=& - \mathcal{N} g_1 \re^{\ri k_3 x}, \nn \\
  f_2 &=& 0.
\eea
with the normalisation factor $\mathcal{N} = (2\pi(g_1^2+g_3^2))^{-1/2}$.
The wavenumbers are related to the eigenenergy by $ck_1 = E - E_1$ and 
$ck_3 = E - E_3$.

The other solution is discontinuous due to the interaction with the atom.
One finds that
\bea
  f_1(x) &=& \mathcal{N} g_1 
  \left[ \Theta(-x) + t_E \Theta(x) \right]  \re^{\ri k_1 x}, \nn \\
  f_3(x) &=& \mathcal{N} g_3
  \left[ \Theta(-x) + t_E \Theta(x) \right] \re^{\ri k_3 x}, \nn \\
  f_2 &=& \ri c (t_k -1),
\eea
with the transmission coefficient
\be
  t_E = \frac{(E - E_2+\ri \gamma/2) - \ri(\Gamma_1 + \Gamma_3)/2}{
    (E - E_2 + \ri\gamma/2) + \ri (\Gamma_1 + \Gamma_3)/2},
\ee
where $\Gamma_j = g_j^2/2$ are the spontaneous emission rates
into the waveguide when the emitter decays into the respective 
atomic level.

Using these states, one can easily solve the scattering problem.
We find that an incoming photon with wavenumber $k$ is scattered 
depending on the atomic state according to
\bea
  \ket{k,1}_e &\rightarrow& \frac{t_E \Gamma_1+\Gamma_3}{
                   \Gamma_1+\Gamma_3}  \ket{k,1}_e
            + \frac{\sqrt{\Gamma_1 \Gamma_3} (t_E - 1)}{\Gamma_1+\Gamma_3}  
                 \ket{k - q, 3}_e \nn \\
		&& \mbox{with} \; E = E_1 + ck, \nn \\ [2mm]
  \ket{k,3}_e &\rightarrow& \frac{\Gamma_1 + t_E \Gamma_3}{
                  \Gamma_1+\Gamma_3} \ket{k,3}_e
            + \frac{\sqrt{\Gamma_1 \Gamma_3}  (t_E - 1)}{
                 \Gamma_1+\Gamma_3}  \ket{k + q, 1}_e \nn \\
		&& \mbox{with} \; E = E_3 + ck,
		\label{eqn-stot-l2}
\eea
where $cq = (E_3 -E_1)$ is the energy difference between the two ground 
states.

A remarkable result is that a resonant photon can be absorbed deterministically
as pointed out in \cite{Pino08} for the case $\gamma = 0$.
Suppose the emitter is initially prepared in the ground state $\ket{1}$ and the 
coupling constants satisfy the condition $\Gamma_1^2-\Gamma_3^2 =  
\gamma(\Gamma_1 + \Gamma_3)$. The transmission amplitude is then 
given by $t_E = -\Gamma_3/\Gamma_1$ and the internal state of the emitter 
is flipped,
\be
  \ket{k,1}_e \rightarrow - \sqrt{\Gamma_3/\Gamma_1} \; \ket{k-q,3}_e,
\ee
as long as the photon is not lost. If transversal losses are absent, i.e. 
$\gamma = 0$, the scattering process realizes a deterministic quantum
gate between a photon and a single emitter. However, the described 
protocol is also very useful in the case $\gamma \neq 0$, when a 
significant fraction of the incident photons is lost. If a single photon 
is measured in the output, the gate is known to have operated successfully.
Combined with error-proof quantum communication schemes
\cite{Enk97,Sore98,08bell}, this provides a powerful building
block for quantum computing purposes. We will discuss 
the realization of a single photon transistor using this property
in section \ref{sec-transistor}.

\subsection{Driven V-type atom}
\label{sec-driven-v}

Next we consider the situation shown in figure \ref{fig-levels3} (C), where the 
classical driving field couples the ground state $\ket{1}$ to the excited state 
$\ket{3}$, so that the atomic Hamiltonian is now given by
\bea
  H_{\rm atom} =  (E_2- \ri \gamma/2) \ket{2}\bra{2} - \Delta \ket{3}\bra{3} 
   + \frac{\Omega}{2} \left( \ket{3}\bra{1} + \ket{1}\ket{3} \right).
\eea
The interaction Hamiltonian is given by equation (\ref{eqn-ham1}) with
$\hat S_+ = \ket{2}\bra{1}$. The state $\ket{3}$ is assumed to be stable,
for instance the states $\ket{1}$ and $\ket{3}$ can be two hyperfine levels 
coupled by microwave radiation.

The situation is best dealt with in a dressed state picture, introducing
the eigenstates of the atomic Hamiltonian
\bea
  \ket{+} &=&  \frac{1}{\sqrt{2\Omega_{\rm eff} (\Omega_{\rm eff}-\Delta) }}
  \left( \Omega \ket{1}  + (\Omega_{\rm eff} - \Delta) \ket{3}   \right)          \nn \\
  \ket{-} &=& \frac{1}{\sqrt{2\Omega_{\rm eff} (\Omega_{\rm eff}+\Delta) }}
    \left( \Omega \ket{1}  - (\Omega_{\rm eff} + \Delta) \ket{3}   \right)  \nn \\
\eea
with eigenenergies
\be
  E_\pm = - \frac{\Delta}{2} \pm \frac{\Omega_{\rm eff}}{2}. 
\ee
The atomic and interaction Hamiltonians are then given by
\bea
  H_{\rm atom} &=& E_+ \ket{+}\bra{+} +  E_- \ket{-}\bra{-}  
    + (E_2-i\gamma/2)  \ket{2}\bra{2}  \nn \\
  H_{\rm int} &=& \int \rd x \, \delta(x) \, \hat a^\dagger(x) 
   \big[g_+  \ket{+}\bra{2} 
      + g_- \hat a^\dagger(x) 
      \ket{-}\bra{2} + \mbox{h.c.} \big] \, .
\eea
with
\bea
  g_\pm = \frac{g \Omega}{
     \sqrt{ 2 \Omega_{\rm eff} (\Omega_{\rm eff} \mp \Delta)   } } .
\eea
This is exactly the situation depicted in figure \ref{fig-levels3} (B), 
which has been discussed in the previous section. However, a 
driven system has the enormous advantage that system 
parameters such as the ration $g_+/g_-$ can be tuned to 
almost any desired value by choosing an appropriate classical 
control field. In particular, this can be used to implement a single 
photon transistor as shown in section \ref{sec-transistor}.

Using the result from the previous section, one finds that an 
incoming photon with wavenumber $k$ is scattered according to
\bea
  \ket{k,+}_e &\rightarrow& \frac{t_E \Gamma_+ + \Gamma_-}{
                      \Gamma_+ + \Gamma_-} \ket{k,+}_e 
                                  + \frac{\sqrt{\Gamma_+ \Gamma_-} (t_E - 1)}{ 
                     \Gamma_+ + \Gamma_-}  \ket{k - q, -}_e \nn \\
		&& \qquad \mbox{with} \; E = E_+ + ck - E_2 \nn \\ [2mm]
  \ket{k,-}_e &\rightarrow& \frac{ \Gamma_+ + t_E \Gamma_-}{
                 \Gamma_+ + \Gamma_-} \ket{k,-}_e
                  + \frac{\sqrt{\Gamma_+ \Gamma_-}  (t_E - 1)}{ 
                 \Gamma_+ + \Gamma_-}  \ket{k + q, +}_e \nn \\
		&& \qquad \mbox{with} \; E = E_- + ck - E_2, 
		\label{eqn-dv-scat}
\eea
where $cq = (E_- -E_+)$ is the energy difference between the two 
dressed states. The transmission phase factor is given by
\be
  t_E = \frac{(E  -E_2 + \ri \gamma/2 ) - \ri (\Gamma_+ + \Gamma_-)/2}{
   (E - E_2 + \ri \gamma/2)  + \ri (\Gamma_+ + \Gamma_-)/2}.
\ee

\subsection{A V-type atom with two coupling transitions}

This case, depicted in figure \ref{fig-levels3} (D), is equivalent to 
case A in the dressed state picture as long as loss can be neglected.
In the general case one has to be a bit more careful, because the loss
terms are also changed by a transformation to the dressed state
basis. Thus we also give the full solution for the case of a V-type
atom with two coupling transitions in the following.

We consider the atomic Hamiltonian
\be
  H_{\rm atom} =
        (E_2 - \ri \gamma_2/2) \ket{2}\bra{2} 
          + (E_3 - \ri \gamma_3/2 ) \ket{3}\bra{3} 
\ee
where we have set the energy scale such that the energy of level $\ket{1}$
is zero. The interaction Hamiltonian is given by 
\be
  \hat H_{\rm int} = \int \rd x \, \delta(x) \;
  \left(
   g_2  \hat a_e(x) \ket{2} \bra{1} + g_3 \hat a_e(x) \ket{3} \bra{1} + \mbox{h.c.}
   \right)
\ee
The eigenstates of the full Hamiltonian can then be written as
\be
  \ket{E} = \int \rd x f_1(x) \hat a^\dagger(x) \ket{\emptyset,1}
     + f_2 \ket{\emptyset,2} + f_3 \ket{\emptyset,3}
\ee
with eigenenergy $E = ck$. The coefficients are easily found to be
\bea
  f_1(x) &=& \frac{1}{\sqrt{2\pi}} \left( \Theta(-x) + t_k \Theta(+x) \right) 
              \re^{\ri kx} \nn \\
  f_2 &=& \frac{1}{2\sqrt{2\pi}} \, \frac{\sqrt{c\Gamma_2} (1+t_k)}{
            ck - E_2 + \ri \gamma_2/2}	\nn \\
  f_3 &=&  \frac{1}{2\sqrt{2\pi}} \, \frac{\sqrt{c\Gamma_3} (1+t_k)}{
                        ck - E_3 + \ri \gamma_3/2} \nn,
\eea
where the transmission coefficient is given by
\be \fl \qquad 
  t_k = \frac{[ck-(E_2-\ri \gamma_2/2) - \ri \Gamma_2/2 ]
                   [ck-(E_3-\ri \gamma_3/2) - \ri \Gamma_3/2] 
                          - \Gamma_2 \Gamma_3 /4}{
		   [ck-(E_2-\ri \gamma_2/2) + \ri \Gamma_2/2 ]
                   [ck-(E_3-\ri \gamma_3/2) + \ri \Gamma_3/2] 
                          - \Gamma_2 \Gamma_3 /4} \, .
\ee
An incoming photon with wavenumber $k$ in the $e$-mode thus 
experiences a phase shift given by $t_k$ when it crosses the atom. 
The reflection and transmission amplitudes for the left- and right-moving
are again given by equation (\ref{eqn-rt-tilde}). 
The atom can become fully transparent only if one of the levels 
$\ket{2}$ or $\ket{3}$ is metastable and the photon is resonant to
the atomic transition between the metastable and the ground level:
\be
   |\tilde t_k| = 1 \quad \mbox{for}  \quad 
        \left\{ \begin{array}{l}
         ck = E_2 \; \mbox{and} \; \gamma_2 = 0 \qquad \mbox{or} \\
   	 ck = E_3 \; \mbox{and} \; \gamma_3 = 0. \\ 
	 \end{array} \right. 
\ee

\section{Applications}

Having solved the basic scattering problem, we now go on to discuss
various applications of these results.

\subsection{A single photon transistor}
\label{sec-transistor}

A single photon transistor is a universal nonlinear element with
wide-ranging application from classical telecommunication to
quantum computing \cite{Bouw00,Chan07b}.
However, it is very hard to realize since photons rarely interact 
directly, so that the necessary interactions must be mediated by 
matter in a deterministic way.

The basic idea of this transistor is that a waveguide strongly coupled 
to a single emitter is either reflective or transparent, depending on the 
state of the emitter. The state can then be switched by a single gate photon.
Previously, a realization of such a device in nanoscale one-dimensional
waveguides has been proposed \cite{Chan07b}. A disadvantage of this 
method is that it requires a precise timing between the quantum gate 
photon and a classical control pulse. Here we present an alternative 
approach which removes this restriction.

This single photon transistor works as follows: 
Consider the situation sketched in figure \ref{fig-levels3} (C), whose 
scattering properties have been analyzed in section \ref{sec-driven-v}.
Initially the system is in state $\ket{1}$ and the classical control field
is off $\Omega(-\infty) = 0$. 
Now the emitter is driven to the state $\ket{+}$ by an adiabatic passage
technique, slowly increasing $\Omega$ and possibly varying the 
Laser frequency so that finally a situation with $\Omega \neq 0$ and
$\Delta=0$ is reached. Several methods for this procedure
were demonstrated experimentally (see, e.g., \cite{Ober06}).
After this initialisation procedure, the transistor is sensitive 
to the arrival of the gate photon as long as the driving field remains
on. If one is injected, it scatters 
according to equation (\ref{eqn-dv-scat}). As we have prepared the emitter 
such that $\Delta = 0$ we have $\Gamma_+ = \Gamma_- 
= \Gamma/\sqrt{2} $. Assuming 
that the photon is resonant,  $ck = E_2 - E_+$, the transmission
coefficient is given by $t_E = -1$ as long as losses can be neglected.
Thus the internal state of the emitter is changed during the
scattering process according to
\begin{eqnarray}
   && \mbox{gate photon present:} \; \ket{k,+} \rightarrow - \ket{k,-} \nn \\
   && \mbox{no gate photon:} \; \ket{k,+} \rightarrow  \ket{k,+}. \nn
\end{eqnarray}
After the scattering process, the initialisation sequence is reversed,
mapping $\ket{+}$ back to $\ket{1}$ and $\ket{-}$ to $\ket{3}$.
Thus the state of the emitter is switched from $\ket{1}$ to $\ket{3}$
if and only if a gate photon has crossed the waveguide.
As the external field is now switched off, $\Omega = 0$, the scattering
is essentially that of a two-level emitter. As shown in \cite{Shen05}, the
waveguide is reflective if the emitter is in state $\ket{1}$. If it has been
switched to state $\ket{3}$ by the gate photon, it is essentially decoupled
-- the waveguide is transparent. 
It should be noted that this procedure does not need a precise
timing. The initialisation and de-initialisation sequences must be 
performed well before and well after the gate photon passed but
the precise time does not play a role.
Furthermore, the de-initialisation is performed by an adiabatic 
passage, which is more robust than procedures relying on Rabi 
flopping.

\begin{figure}[tb]
\centering
\includegraphics[width=8cm, angle=0]{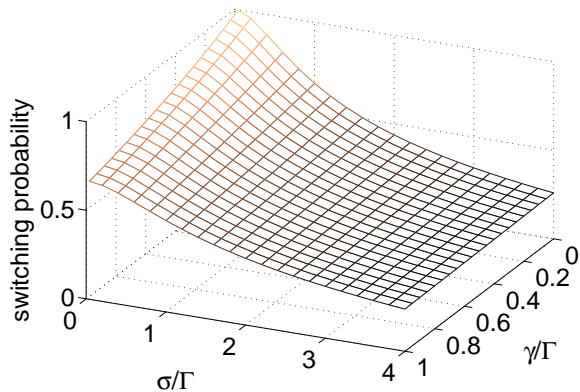}
\caption{\label{fig-transistor-ps}
Switching probability of the single photon transistor described
in the text as a function of the loss rate $\gamma$ and the 
frequency width of the incident photon $\sigma$
for $\Delta= 0$.
}
\end{figure}

In the description of the basic idea we neglected losses and assumed
a perfect resonance. Let us now consider a more realistic situation
and calculate the switching probability for a general input state
\be
  \ket{\Psi_{\rm in}} = \int \rd k \, f(k) \, a_k^\dagger \ket{+,\emptyset}.
\ee  
In particular we consider a Gaussian input pulse
\bea
     f(k) = \frac{1}{\sqrt[4]{2\pi\sigma^2}} 
      \re^{-(ck-\omega_0)^2/4\sigma^2},
    \label{eqn-gauss}
\eea
at resonance with the transition $ \omega_0 = E_2 - E_+ $.
Assuming that the adiabatic passage to the dressed state 
works perfectly the switching probability is then given by the 
probability to find the emitter in state $\ket{-}$ after the 
scattering:
\be
  p_s = \int  | f(k)|^2 \frac{|t_E - 1|^2}{4} \ \rd k  \, + \,
            \int  | f(k)|^2 \frac{1- |t_E|^2}{4} \, \rd k.
\ee
The first term gives the norm of the contribution of $\ket{-}$ 
in the scattering state (\ref{eqn-dv-scat}), weighted by the pulse 
shape of the gate photon and using $\Gamma_+ = \Gamma_-$.
The second contribution is the probability that the emitter decays 
to level $\ket{-}$ when the gate photon is lost by spontaneous 
emission to the outside world, where the branching ratio is $1/2$.
This switching probability is plotted in the case $\Delta = 0$ as a 
function of the width of the input pulse $\sigma$ and the loss rate 
$\gamma$ in figure \ref{fig-transistor-ps}.

\subsection{Tunable photonic band gaps}
\label{sec-bloch}

\begin{figure}[tb]
\centering
\includegraphics[width=6cm, angle=0]{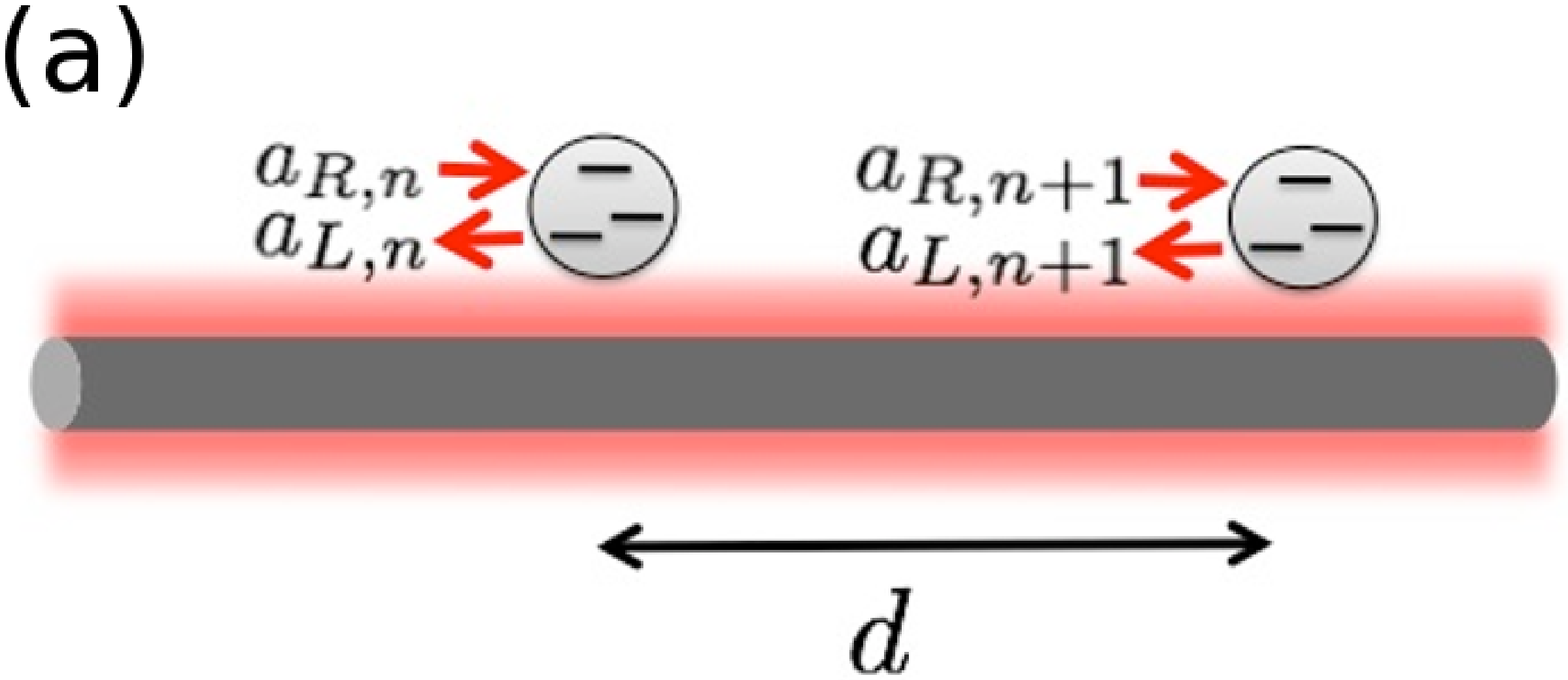}
\includegraphics[width=8cm, angle=0]{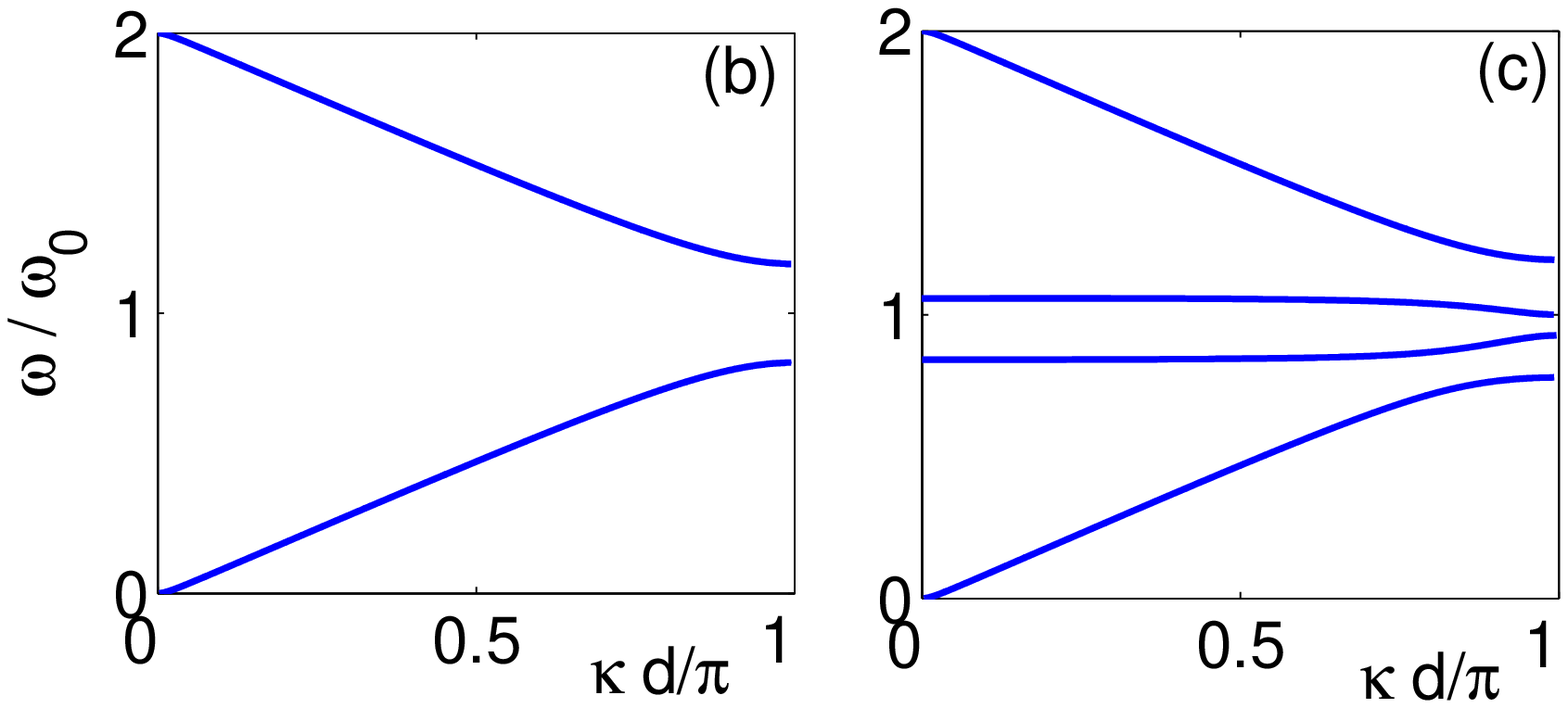}
\caption{\label{fig-array}
(a) Waveguide coupled to a periodic array of emitter of emitters 
in the EIT configuration (cf. figure \ref{fig-levels3} A).
(b) Photonic Bloch bands of the waveguide without driving for
$\Gamma = 0.1 \omega_0$ and $d = 0.5 \lambda_0$, 
where  $\omega_0 = (E_2 - E_1)$ is the resonance 
frequency and $\lambda_0 = 2\pi c / \omega_0$.
(c) Driving the emitters with a Rabi frequency of 
$\Omega/\omega_0 = 0.2$ and $\Delta = 0.1 \, \omega_0$
leads to a splitting of the states around the resonance
frequency and the occurence of new subbands.
}
\end{figure}

Let us now consider the properties of a periodic array of emitters 
as shown in figure \ref{fig-array}. The emitters are driven by a classical 
laser field in the EIT configuration shown in figure \ref{fig-levels3} (A).
The strength and detuning of this driving then allows to tune
the transport properties of the array.

Let us first consider the lossless case $\gamma_2 = \gamma_3 = 0$.
The amplitudes of the right- and left- going waves in this array
are then related by
\be
     \left(\begin{array}{c} a_{R,n+1} \\ a_{L,n+1}  \end{array} \right)  
    =  T  \left(\begin{array}{c c} a_{R,n} \\ a_{L,n}  \end{array} \right)  
\ee
where 
\be
  T = \left(\begin{array}{c c}
  \re^{\ri \omega d/c} & 0 \\ 0 & \re^{-\ri \omega d/c}
   \end{array} \right)  
  \left(\begin{array}{c c}
  1/\tilde t^* & - \tilde r^*/\tilde t^* \\ 
  - \tilde r/\tilde t & 1/\tilde t
  \end{array} \right) 
  \label{eqn-tmatrix}
\ee
is the transfer matrix (cf. \cite{Berr97}) 
Here, $d$ denotes the periodicity of the 
emitter array and the transmission and reflection coefficients 
defined in equation (\ref{eqn-rt-tilde}) have been used.
The eigenvalues of the transfer matrix characterise 
the transport properties of the array. Since the product of both
eigenvalues is always one,  the eigenvalues are either complex 
conjugate numbers of magnitude one ($e^{\pm i\kappa d}$) or they 
are inverse of each other with magnitudes smaller and larger than one. 
The first case characterizes a photonic Bloch state with quasi
momentum $\kappa$, so that transport is possible.
In the latter case transport is impossible, i.e. $\omega$ lies
in a band gap.

Examples of the resulting photonic Bloch bands $\omega(\kappa)$ are 
shown in figure \ref{fig-array}. For $\Omega = 0$ (no driving) one recovers
the Bloch bands well known for simple two-level emitters \cite{Shen05},
with a band gap at the resonance frequency.
For $\Omega >0$ however, the upper level splits into two dressed
states which both give rise to a Bloch band. The position and width
of these bands is then tunable through $\Omega$ and $\Delta$. 

\begin{figure}[tb]
\centering
\includegraphics[width=14cm, angle=0]{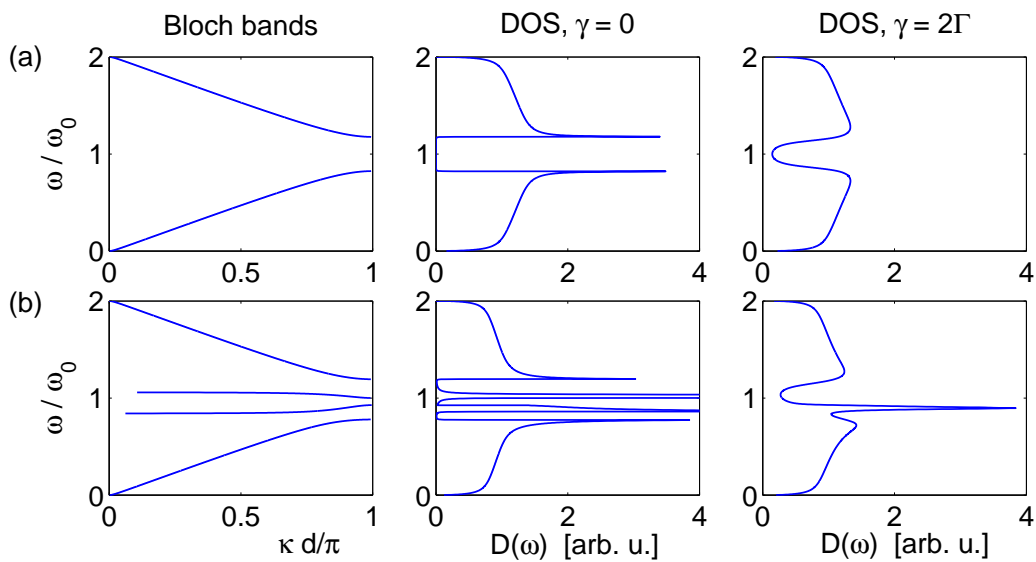}
\caption{\label{fig-array-loss}
Photonic Bloch bands for the same parameters as in figure 
\ref{fig-array} without (a) and with (b) external driving field.
The driving leads to a splitting of Bloch states around the resonance
frequency and thus leads to a drastic modification of the density
of states. These features are still very pronounces in the case 
of strong transversal losses, i.e. large values of $\gamma$.
}
\end{figure}

In a realistic system, the coupling to the waveguide will not
be perfect and transversal losses by spontaneous decay from 
the excited state $\ket{2}$ cannot be neglected.
Then the transfer matrix is no longer unitary and its eigenvalues
are rather given by $\re^{\sigma d} \re^{\pm \ri \kappa d}$, where 
$\sigma$ gives the absorption coefficient.
The effect of losses on the photonic band gap structure is illustrated
in figure \ref{fig-array-loss}, where we compare the density of states
$D(\omega)$ with and without transversal losses.
In the lossless case, the density of states $D(\omega)$
is essentially given by $(\rd\omega/\rd \kappa)^{-1}$. 
For $\gamma \neq 0$, every state is broadened, so that
the density of states is rather given by
\bea
   D(\omega') &=& \int \rd \kappa \, |\chi(\kappa)|^2
      \frac{\sigma(\kappa)/2\pi}{
      (\omega(\kappa)-\omega')^2 + \sigma{\kappa}^2} \nn \\
     &=& \int \rd \omega  \, |\chi(\omega)|^2
     \left( \frac{\rd \omega}{\rd \kappa} \right)^{-1}
        \frac{\sigma(\omega)/2\pi}{
      (\omega-\omega')^2 + \sigma(\omega)^2} .
\eea
This definition of the density of states is adapted to the modification
of the spontaneous emission rate of a single emitter in a photonic 
crystal structure \cite{Yabl87}. Here, $\chi(\kappa)$ denotes the 
coupling matrix element of the emitter to a Bloch state with quasi 
momentum $\kappa$. As the interaction with the waveguide
is local at the position $x_0$ of the emitter (cf.~equation 
(\ref{eqn-ham-int})), the coupling is simply given by the magnitude 
of the Bloch wave function at the position $x_0$:
\be
  \chi(\kappa) =  a_{R,\kappa} \re^{\ri \omega x_0/c}
                           + a_{L,\kappa} \re^{-\ri \omega x_0/c}.
\ee
In the following example we chose $x_0 = d/2$, i.e. we consider
the emission of a single impurity atom placed in the middle 
between two emitters forming the photonic band gap structure. 

Figure \ref{fig-array-loss} shows that the density of states around 
the resonance frequency is strongly modified around the resonance 
frequency $\omega_0$ by the external driving field. 
These features are still very pronounces in 
the case of strong transversal losses, i.e. large values of $\gamma$.
This can be utilized, for instance, in order to manipulate the emission 
rate of a single impurity atom coupled to the waveguide simply by 
tuning the strength $\Omega$ of the external driving laser
(cf. \cite{Yabl87,Loda04}).

\subsection{Tunable Anderson localization of photons}
\label{sec-anderson}

In the previous section we considered transport of photons through
a waveguide coupled to an array of emitters in a perfectly periodic 
configuration. If this periodicity is disturbed by disorder, as  it will 
commonly be in real experiments, there are generally no infinitely 
extended Bloch states any more. Instead, every photonic wavefunction 
undergoes Anderson localization \cite{Ande58}. This localization
was demonstrated experimentally in different systems 
\cite{Wier97,Schw07}.
The localization transition is well understood for systems described by
transfer matrices \cite{Berr97} such as  (\ref{eqn-tmatrix}).  Here we 
consider an array of emitters where the distance between the emitters
$d_n$ differs from site to site, in particular assuming a uniform
distribution in the interval $[0.4,0.6] \times \lambda_0$, where 
$\lambda_0$ is the free-space wavelength of a photon with frequency 
$\omega_0$. The transfer matrix of a whole array of $N$ emitters is 
then given by the product $T_{N} = \prod_{j=1}^N T_j$ and its eigenvalues 
can be written as $\re^{\pm N d_0/\xi}$. The localization length $\xi$ 
characterises the spatial extend of a photonic wavefunction and 
measures the localization strength.

\begin{figure}[tb]
\centering
\includegraphics[width=14cm, angle=0]{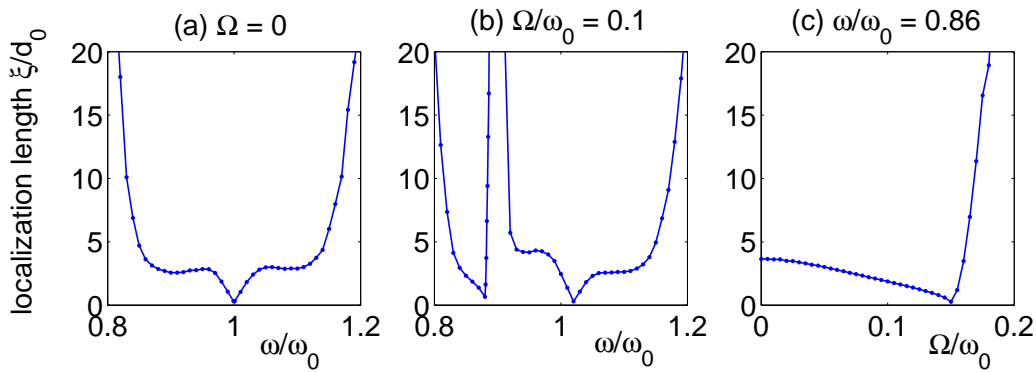}
\caption{\label{fig-anderson}
Photonic Anderson localization in a waveguide strongly coupled
to an array of emitters at random positions.
(a) Localization length for an array of three-level emitters with 
parameters $\Gamma = 0.1 \omega_0$ and average emitter 
distance $\bar d = 0.5 \lambda_0$, where
$\omega_0 = (E_2 - E_1)$ is the resonance frequency
and $\lambda_0 = 2\pi c / \omega_0$.
(b) Driving the emitter with the Rabi frequency  
$\Omega/\omega_0 = 0.1$ and $\Delta/\omega_0 = 0.1$ 
alters the localization properties fundamentally.
(c) Localization length as a function of the strength of the 
external driving field for $\omega/\omega_0 = 0.86$
}
\end{figure}

An example is shown in figure \ref{fig-anderson} for an array of 
three-level emitters in the EIT configuration as shown in 
figure \ref{fig-levels3} (A). We restrict ourselves to the lossless
case and set $\Gamma = 0.1 \omega_0$, where 
$\omega_0 = (E_2 - E_1)$ is the atomic resonance frequency.
The localization length $\xi$ is plotted as a function of the 
frequency $\omega$ of the incoming photons as obtained
from a Monte Carlo simulation. The transfer matrix $T_N$ has
been computed using $N=100$ randomly chosen values for
the distance $d_n$ and diagonalized. The obtained values  
for the localization constant $\xi^{-1}$ have been averaged over 100
realizations.
Without external driving (figure \ref{fig-anderson} a), one observes 
the localization spectrum of a single two-level
emitter. On resonance all emitters are completely reflective so that
the localization length tends to zero. The introduction of a classical
driving field changes the localization properties dramatically
(cf. figure \ref{fig-anderson} b).
The upper state splits into two dressed states, so that there are two 
resonance frequencies for which the emitters are fully reflective and 
hence $\xi  = 0$. On the other hand $\xi$ becomes infinite when the
EIT condition is fulfilled, as all emitter are fully transparent then.

\section{Conclusion and Outlook}

In the present paper we have solved the scattering problem for
a single photon in a one-dimensional waveguide coupled to a
three-level emitter. Several different configurations were taken
into acount. Electromagnetically induced tranparency is observed
for a driven $\Lambda$-system and $V$-system if both transitions
couple to the waveguide. On the contrary, scattering to sidebands
occurs for  a driven $V$-system and $\Lambda$-system with two 
coupling transitions.

The control gained by the classical driving field pave the way for a
variety of application from classical optics to quantum information.
The control over the transmission in the EIT scheme can be used to
taylor the optical properties of an array of emitters. Thus one can
readily engineer photonic band gap structures or tune the localization
length in disordered systems.
The driven $V$-system can be applied in a single photon transistor.
The gate photon is scattered to a sideband changing the internal
state of the emitter and thus switching the waveguide from reflective
to transmittive.

\section*{Acknowledgements}

Financial support from the Deutsche Forschungsgemeinschaft via the research
fellowship programme (grant number WI 3415/1) and the Villum Kann Rasmussen
foundation is gratefully acknowledged.

\section*{References}

\providecommand{\newblock}{}

\end{document}